\documentclass[fleqn,12pt,twoside]{article}
\usepackage{amsmath}
\usepackage{espcrc1}

\usepackage{graphicx}
\usepackage[figuresright]{rotating}

\newcommand{\figs}{}
\title{Flavor independent systematics of excited
baryons and intra-band transition%
\footnote{Presented by M.~Koma at ``PaNic02'',
Osaka, Japan, 30 Sep. - 4 Oct. 2002}   }                            

\author{M. Koma (Takayama),\address[MPI]{
Max-Planck-Institut f\"ur Physik, F\"ohringer Ring 6, D-80805
M\"unchen, Germany \\ 
}\address[RCNP]{Research Center for Nuclear Physics, 
Osaka University, 
Osaka 567-0047,
Japan}
A.~Hosaka\addressmark[RCNP]
and
H.~Toki \addressmark[RCNP]
}       
\begin{document}

\maketitle

\begin{abstract}
    Transitions among excited nucleons are studied within a
    non-relativistic quark model with a deformed harmonic oscillator
    potential.
    The transition amplitudes are factorized into the $l$-th moment and a
    geometrical factor. This fact leads to an analogous result to the
    ``Alaga-rule'' for baryons.
\end{abstract}

\section{Introduction}
There are many phenomenological models for 
masses and/or transition properties of baryons.
However, if one tries to extend a model from the flavor SU(2) to SU(3)
or to include excited state, the situation of fitting the experimental data
becomes more and more complicated.  Even if the spectrum can be
reproduced by a phenomenological model, it is not easy to extract
underlying physics or intuitive picture from them
since some of the important effects may be hidden
behind the model parameters.  This is due to the seemingly complicated
structure of the experimental data.
It seems important therefore to have a fresh look at the existing
data.

Historically, it is known that the Gell-Mann-Okubo (GO) mass 
formula works well to describe the pattern of the flavor 
octet and decuplet baryons in the ground states.
We have extended this GO mass formula to all observed
flavor SU(3) baryons 
and found that there is a flavor independent systematics in
spectra~\cite{DOQsys,takayama-panic-1999}, more precisely in the
excitation energy.  
Remarkably, we have found that 
these patterns of mass spectra have  a  quite similar behavior
to the {\it rotational band} in the deformed nuclei, which seem to imply
spatial deformation of excited states of baryons.  We have then
investigated these spectra in terms of an effective non-relativistic
quark model with a deformed harmonic oscillator potential, which we
call the deformed oscillator quark (DOQ) model, and have shown that almost
all data can be reproduced with only one parameter by the DOQ model.

\par
In the deformed nuclei,  typical transitions are observed
among rotational bands, which are classified as 
{\it inter}-band  
or
{\it intra}-band transitions.
We expect that  we can also observe
similar transitions in the excited baryons if the
pattern of their spectra is described by rotational bands.
In this brief report, we study intra-band 
transitions  which correspond to the transitions among 
excited baryons with the same parity through an emission of a pion
in the framework of the DOQ model.

\section{The DOQ model}
Let us start with the hamiltonian of the DOQ model:
\begin{align} H_{\mathrm{DOQ}}  & = 
    \sum_{i=1}^{3} 
    \left(
     \frac{\vec{p}^{2}_{i}}{2m} + 
    \frac{1}{2}m( \omega^{2}_{x} x^{2}_{i}
    +\omega^{2}_{y}y^{2}_{i} +
    \omega^{2}_{z} z^{2}_{i}
    )   
    \right),
\end{align}
where \( \vec{p}_{i} \) and \( \vec{r}_{i}=(x_{i},y_{i},z_{i}) \) are
the momentum and position of the \( i \)-th quark.  The constituent
quark mass $m$ is taken to be 300 MeV, which is insensitive to the
following calculations.  In order to take into account the deformation
of excited states, the oscillator parameters \( \omega_{x,y,z} \) can
take different values, which are determined by energy minimization \(
\partial E / \partial \omega_{x} = 0, \partial E / \partial \omega_{z}
= 0, \) with volume conservation condition \(
\omega_{x}\omega_{y}\omega_{z} \equiv \omega^{3} = (\mathit{const.})
\), where \( E \) is the eigenenergy of \( H_{\mathrm{DOQ}} \).  The
only one parameter to be fitted in the DOQ model is $\omega$, which
determines the energy scale of the system.  We choose \( \omega = \) 644
MeV to reproduce the excitation energy of the first $1/2^{+}$
excited states.  The system is characterized by the total principle
quantum number \( N = n_{x} + n_{y} + n_{z} \), where \( n \)'s are
oscillator quanta along each axis.  Here we take \( N=n_{z},\, n_{x} =
n_{y} = 0\), since this combination gives the lowest energy for given \( N
\).  Resulting system is spherical for \( N=0 \) (ground state) and 
prolately deformed: 1:2 for \( N=1 \) (excited states with negative
parity) 1:3 for \( N=2 \) (those with positive parity).

Inspired by such a large deformation of the system,
we adopt semi-classical picture to obtain the orbital wave
function with a definite angular momentum \( L \)
and its \( z\)-component \(L_{z} \);
we introduce collective motion of strongly deformed intrinsic state
by Wigner \( D \) function.
Together with the appropriately symmetrized
spin and isospin wave function \( \chi \) and \( \phi \), 
the DOQ wave function with total spin $J$ and its \( z \)-component
\(M \) 
is written as $ \Psi^{N;JM}(\vec{r}) = 
\psi^{JM}(\vec{\Theta})\varphi^{N}_{\mathrm{int}}(\vec{r}_{\mathrm{int}}) 
\,,
 $ where $\psi^{JM}\equiv [D^{L}_{0L_{z}}(\vec{\Theta})
  \chi ]_{JM}$.
Here the collective coordinate and internal body fixed
coordinate are denoted as $\vec{\Theta} $ and $\vec{r}_{\mathrm{int}}$.

\section{Intra-band transitions}
\sloppy{
We calculate transition amplitudes between two excited nucleons
($N^{*}$) with 
a pion,
 $T_{N^{*}(JM) \to \pi N^{*}(J'M')} 
$.}
For \( H_{\pi qq} \),
we take the standard non-relativistic 
\( \pi qq \) interaction,
\begin{align}
    H_{\pi qq}(\vec{r})  & = - \frac{g}{2m}\frac{1}{\sqrt{2\omega_{k}}}
    \left( \vec{\sigma}\cdot\vec{\nabla} \exp (i \vec{k} \cdot
    \vec{r})\right)
    \tau  \,
\end{align}
where \( k \) and \( \omega_{k} \) are the momentum and energy of an
emitted pion, respectively.  Depending on the charge of an emitted
pion, an appropriate component of the Pauli matrix for isospin $\tau$ acts
on the isospin wave function of the $N^{*}$ state. 

Here we adopt a semi-classical cranking method to calculate the
transition amplitudes;
\begin{align} 
    &T_{N^{*}(JM) \to \pi N^{*}(J'M')} =
    \int_{}^{} d^{3}\Theta  \psi ^{*J'M'} (\vec{\Theta})
     \mathcal{O}_{\mathrm{int}} (\vec{\Theta}) 
  \psi^{JM}(\vec{\Theta}),
  \\
  &  \mathcal{O}_{\mathrm{int}} (\vec{\Theta})  = 
  \int_{}^{} [d^{3}r_{\mathrm{int}}] \,
  \varphi^{*N}_{\mathrm{int}}(\vec{r}_{\mathrm{int}})
  R^{\dagger}(\vec{\Theta})  H_{\pi qq}(\vec{r}) R(\vec{\Theta}) 
  \psi^{N}_{\mathrm{int}}(\vec{r}_{\mathrm{int}}),
\end{align}
where $R(\vec{\Theta})$ is a rotation operator which relates the original
coordinate $\vec{r}$ to the internal body-fixed coordinate
$\vec{r}_{\mathrm{int}}$.
Resulting  transition amplitudes are written 
in terms of the \( l \)-th moment $Q^{(l)}(k)
\equiv \int  [d^{3}r_{\mathrm{int}}] 
\varphi^{*N}_{\mathrm{int}}(\vec{r}_{\mathrm{int}}) 
j_{l}(kr)  Y_{l\,0} (\hat{r})
\varphi^{N}_{\mathrm{int}}(\vec{r}_{\mathrm{int}}) 
 $, 
 and some geometrical factors.

\begin{table}
      \caption{Selection rules for strong intra-band decays $N^{*} \to
   N^{*}\pi $ (upper: positive parity, lower: negative parity) }
   \begin{minipage}[]{6cm}
       \centering
       \includegraphics[width = 3.2cm]{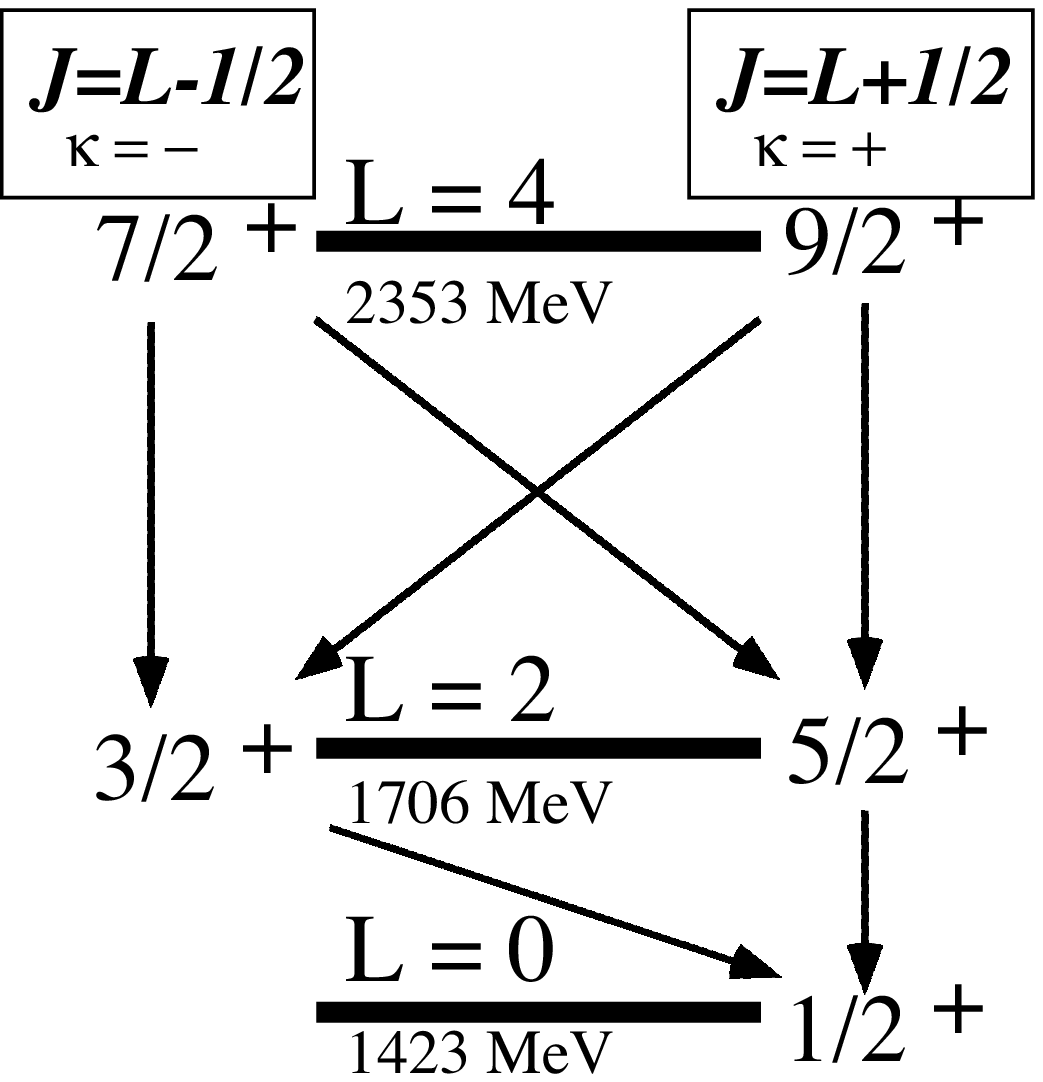}
   \end{minipage}
\begin{tabular}{ccccc}
       \hline
     $L $&\multicolumn{1}{l}{ \( J^{P} \)} &\( J^{\prime P} \) & 
       \(  l_{\pi} \)   &  $\kappa \,\phantom{ \to }
       \,\kappa^{\prime}$\\
       \hline
       &\( 9/2 ^{+}  \to \) & \( 5/2 ^{+} \) & \( 3 \)  & $+ \to +$\\
      $L = 4$& \( 9/2 ^{+}  \to \) & \( 3/2 ^{+} \) & \( 3 \)  & $+ \to -$\\
       &\( 7/2 ^{+}  \to \) & \( 5/2 ^{+} \) & \( 1 \),  \( 3 \) 
       &$- \to+$ \\
       &\( 7/2 ^{+}  \to \) & \( 3/2 ^{+} \) & \( 3^{+} \) & $- \to -$\\
       \hline
	$L = 2$&\( 5/2 ^{+}  \to \) & \( 1/2 ^{+} \) & \( 3 \)  &$+ \to +$\\
       &\( 3/2 ^{+}  \to \) & \( 1/2 ^{+} \) & \( 1 \),  \( 3 \) & $- \to+$\\
       \hline
   \end{tabular}
 
   \begin{minipage}[]{6cm}
       \centering
       \includegraphics[width = 3.2cm]{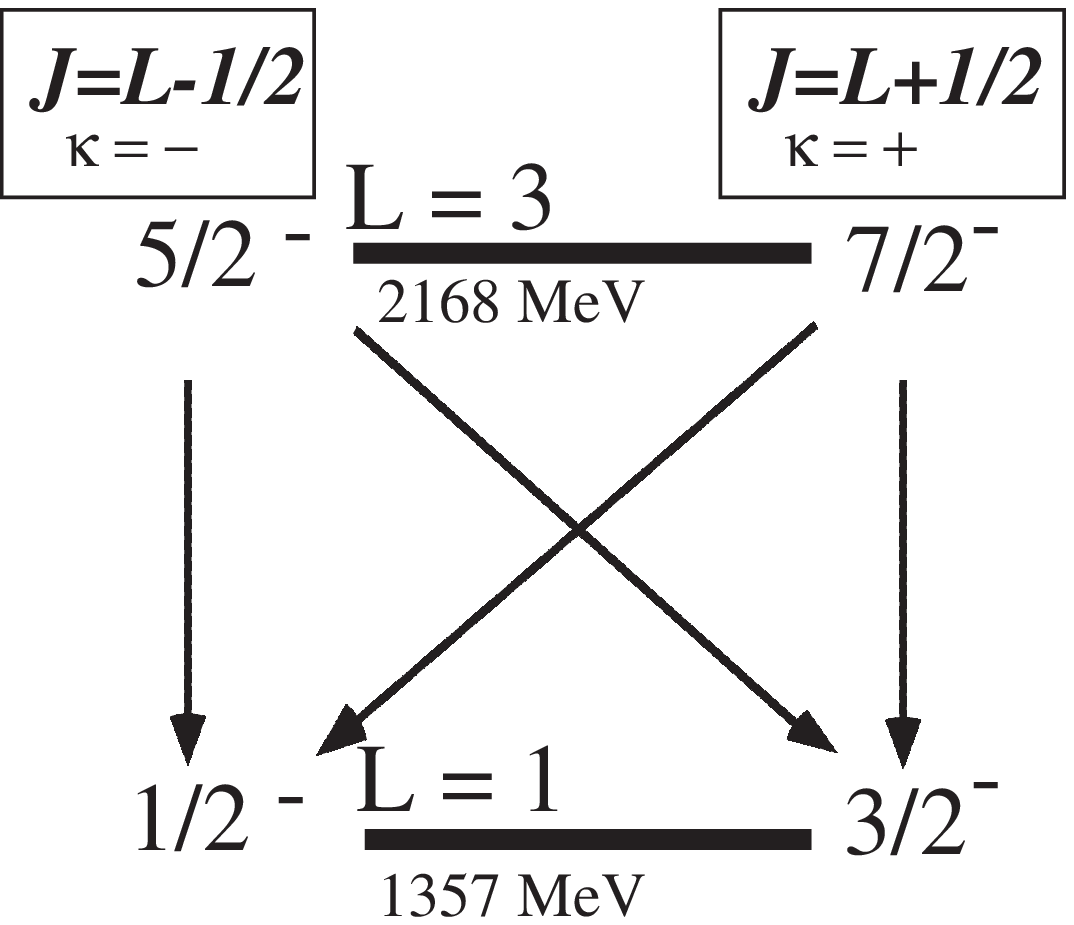}
   \end{minipage}
\begin{tabular}{ccccc}
       \hline
     $L $& \multicolumn{1}{l}{ \( J^{P} \)} &\( J^{\prime P} \) & 
       \(  l_{\pi} \)   &  $\kappa \phantom{ \to } \kappa'$\\
       \hline
       &\( 7/2 ^{-}  \to \) & \( 3/2 ^{-} \) & \( 3 \)  & $+ \to +$\\
      $L = 3$& \( 7/2 ^{-}  \to \) & \( 1/2 ^{-} \) & \( 3 \)  & $+ \to -$\\
       &\( 5/2 ^{-}  \to \) & \( 3/2 ^{-} \) & \( 1 \),  \( 3 \) 
       &$- \to+$ \\
       &\( 5/2 ^{-}  \to \) & \( 1/2 ^{-} \) & \( 3 \) & $- \to -$\\
       \hline
   \end{tabular}
   \label{tbl:negative-pi}
\end{table}

The definition of the decay width $W$ is
\begin{align}
    W & = 6\pi\int_{}^{} \frac{d^{3}k}{(2\pi)^{3}}
    \delta(E'+\omega_k-E)\frac{1}{2J+1}\sum_{MM'}
    \left| T_{N^{*}(JM) \to \pi N^{*}(J'M')} \right|^{2}\,,
    \label{eq:decaydef}
\end{align}
where $E (E')$ is the energy of the initial (final) $N^{*}$ state.
Since both initial and final states are
resonances,  
1) average of the initial spin states, 2) summation over the final spin
states, 3) integration over the momentum of emitted pions and 4)
 an overall factor three for the final state of pions
($\pi^{+},\pi^{-},\pi^{0}$) are taken into account.

Since the difference between angular momentum of initial and final
states is always two for the intra-band transitions,
we can label transitions by $L$, $l_{\pi}$ and
relative orientation of the spin and the orbital angular momentum for
the initial (final) state, $\kappa$ ($\kappa'$).
At the same time, the conservation of the angular momentum
and $p$-wave coupling of pion lead to the selection rules as shown in
Table~\ref{tbl:negative-pi}. 

After some straightforward calculations, 
the decay width \eqref{eq:decaydef} can be written as
\begin{align}
    W& = 	\frac{27 g^{2}}{2m^{2} }	\tilde{k}
    \sum_{l_{\pi}}
    \left|F^{(N)\kappa \kappa'}_{l_{\pi};L}
    ( \tilde{k})\right|^{2},
    \qquad \tilde{k}  = \sqrt{(E_{i} - E_{f})^{2} - m_{\pi}^{2}},
    \label{eq:finalwidth}
\end{align}
where we define transition form factor $F^{(N)\kappa
\kappa'}_{l_{\pi};L}$, where $\kappa(\kappa')$ is either $+$ or $-$ .

\begin{table}[htb]
\caption {Decay widths for each transition are shown in 
    units of MeV. 
    The classification scheme of the column is based on the selection 
    rules given in Table~\ref{tbl:negative-pi} . 
    }
\label{table:1}
\newcommand{\m}{\hphantom{$-$}}
\newcommand{\cc}[1]{\multicolumn{1}{c}{#1}}
\renewcommand{\tabcolsep}{1pc} 
\renewcommand{\arraystretch}{1.2} 
\begin{tabular}{@{}ccccc}
\hline
 $L=4 \to 2$ & \( 9/2^{+} \) $\to$ \( 5/2^{+} \)   
 &\( 9/2^{+} \)  $\to$ \( 3/2^{+} \) 
 &\( 7/2^{+} \)  $\to$ \(  5/2^{+} \) &\( 7/2^{+} \)  $\to$ \( 3/2^{+} \) \\
 & 5.8 & 10.8 & 35.3 & 1.7  \\
  \hline
  $L=2 \to 0$&\( 5/2^{+} \)  $\to$ \( 1/2^{+} \) & - 
  &\( 3/2^{+} \)  $\to$ \(  1/2^{+} \) & - \\
  & 0.1 & - & 0.2 & -  \\
 \hline
  $L=3 \to 1$& \( 7/2^{-} \)  $\to$ \( 3/2^{-} \) 
  & \( 7/2^{-} \)  $\to$ \( 1/2^{-} \) 
  &\( 5/2^{-} \) $\to$ \( 3/2^{-} \) & \( 5/2^{-} \)  $\to$ \( 1/2^{-} \) \\
  & 7.9 & 9.9 & 39.3 & 0.7  \\
 \hline
 \end{tabular}\\[2pt]
 \label{tbl:width-result}
\end{table}

As an example, we present a transition form factor for 
$L-1/2 \to L'+1/2$ transitions, which is dominated by 
$l_{\pi} = 1$ pion:
\begin{align}
     F^{(N)-+}_{l_{\pi} = 3;L}(k)+F^{(N)-+}_{l_{\pi} =1;L}(k)   
    \sim F^{(N)-+}_{l_{\pi} =1;L}(k) =	
    i^{1}(-1)^{L} k
    \frac{\sqrt{L-1}}{ \sqrt{2L-1}}
    \biggl[
    \frac{1}{\sqrt{5}}   Q^{(2)}_{N}(k)
    \biggr]	 \,.
    \label{eq:Fmp1}
\end{align}
From this expression, an interesting relation among transitions is 
obtained;
apart from the effect of momentum transfer $\tilde{k}$,
the ratio of the decay widths is determined only by geometrical 
factors in Eq.~\eqref{eq:Fmp1}
as
\begin{align}
    \frac{W (L\!+\!2 \!-\!1/2 \to L\! +\!1/2)}{W (L\!-\!1/2 \to L\!-\!2\!+\!1/2)} & 
    = \frac{(L+1)(2L-1)}{(2L+3)(L-1)},  &
    \frac{W  (7/2 \to 5/2)}{W(3/2 \to 1/2)} &= \frac{9}{7}.
    \label{eq:alaga-bari}
\end{align}
This relation corresponds to the ``Alaga rule'' in nuclear 
physics~\cite{alaga-1976}.

Numerical results of the decay width for other transitions
are listed in Table\ref{tbl:width-result}.

\section{Summary and discussion}
We have found several interesting features among the intra-band
transitions:
1) the decay widths of
$L\!=\!2 $ to $L\!=\!0$ are very small due to the small momentum 
transfer and hence a small phase space; 
we can show $W$ has large $k$ dependence,  roughly
$W \propto k^{7}$, with the long wave length approximation
($kr \ll 1$).
2) the decay widths of 
$7/2^{+} \!\to\! 5/2^{+}$ and of $5/2^{-} \!\to \!3/2^{-}$
have a rather large value  since
the pion with $l_{\pi} \!=\! 1$ can contribute them
(See, Table~\ref{tbl:negative-pi}).  
Due to the difference of the geometrical factor in the 
transition form factor $F$
the contribution of the $l_{\pi}\!=\!1$ transition is much bigger than that of
$l_{\pi}\!=\!3$ transition. 
3)
the decay widths of $7/2^{+}\! \to\! 3/2^{+}$ and of 
$5/2^{-}\! \to\! 1/2^{-}$ are small again due to the geometrical factor 
in the transition form factor. 
We would like to mention that such a interpretation of the result
is possible by virtue of the semi-classical cranking formula.

Experimentally, such intra-band transitions may be observed in
$N^{*} \to N\pi\pi$ processes. However,  no clear evidence  has been
found in
existing data at the moment. 
As a future work, we would like to study 
other transition properties such as inter-band transitions with a
pion, which can be comparable with experimental data.


\end{document}